\documentclass[aps,prl,twocolumn,groupedaddress,amsmath,nofootinbib]{revtex4-2}
\usepackage{graphicx}
\usepackage{subfigure}
\usepackage{mathtools}
\usepackage{mathrsfs} 
\usepackage{dcolumn}
\usepackage{bm}
\usepackage{amsmath}
\usepackage{color}
\usepackage{verbatim}
\usepackage{natbib} 
\usepackage{verbatim}
\usepackage{amsfonts}
\usepackage{lipsum}
\usepackage{epstopdf}
\usepackage[colorlinks,
            linkcolor=blue,
            anchorcolor=blue,
            citecolor=blue,
urlcolor=blue
            ]{hyperref}
\begin{document}


\title{Achieving the Multi-parameter Quantum Cram\'er-Rao Bound with Antiunitary Symmetry}



\author{Ben Wang$^{1,\ddagger}$}
\author{Kaimin Zheng$^{1,\ddagger}$}
\author{Qian Xie$^{1}$}
\author{Aonan Zhang$^{1}$}
\author{Liang Xu$^{1}$}
\email{liangxu.ceas@nju.edu.cn}
\author{Lijian Zhang$^{1}$}
\email{lijian.zhang@nju.edu.cn}
\affiliation{$^1$ National Laboratory of Solid State Microstructures, Key Laboratory of Intelligent Optical Sensing and Manipulation, College of Engineering and Applied Sciences, and Collaborative Innovation Center of Advanced Microstructures, Nanjing University, Nanjing 210093, China\\
$^\ddagger$ These authors contributed equally to this work\\
}

\date{\today}

\pacs{}
\begin{abstract}
The estimation of multiple parameters is a ubiquitous requirement in many quantum metrology applications. However, achieving the ultimate precision limit, i.e. the quantum Cram\'er-Rao bound, becomes challenging in these scenarios compared to single parameter estimation. To address this issue, optimizing the parameters encoding strategies with the aid of antiunitary symmetry is a novel and comprehensive approach. For demonstration, we propose two types of quantum statistical models exhibiting antiunitary symmetry in experiments. The results showcase the simultaneous achievement of ultimate precision for multiple parameters without any trade-off and the precision is improved at least twice compared to conventional encoding strategies. Our work emphasizes the significant potential of antiunitary symmetry in addressing multi-parameter estimation problems.

\end{abstract}
\maketitle

\textit{Introduction.}--Quantum metrology
holds the promise of achieving better precision of
parameter estimation than that of classical schemes. Single parameter estimation has achieved great success and finds applications in fields including quantum sensing~\cite{tsang2011fundamental,RevModPhys.89.035002}, and gravitational wave detection~\cite{aasi2015advanced,aasi2013enhanced}. However, practical applications often involve multiple parameters.
The quantum-enhanced measurement of multi-parameter estimation is more challenging compared to single parameter scenarios, and several methods are proposed to enhance the precision including
probe state optimization~\cite{PhysRevLett.125.020501,RN38}, sequential optimal control~\cite{doi:10.1126/sciadv.abd2986,PhysRevLett.126.070503,PhysRevLett.117.160801,PhysRevA.96.042114}, and collective measurement~\cite{RN37,hou_deterministic_2018}. Apart from the technical difficulties, the fundamental bound of precision limit in the single parameter case, i.e. the quantum Cramér-Rao bound (QCRB)~\cite{braunstein1994statistical,Liu_2019,doi:10.1142/S0219749909004839}, may not be saturated in multi-parameter estimation. Various types of bounds are proposed to address this issue~\cite{PhysRevLett.126.120503,PhysRevLett.128.250502,holevo2011probabilistic,hayashi2005asymptotic,PhysRevLett.123.200503,PhysRevX.11.011028,demkowicz2020multi,guctua2006local,hayashi2008asymptotic,kahn2009local,yamagata2013quantum}, which provide precisions worse than that given by the QCRB. Moreover, it remains elusive whether these bounds can be achieved and how to saturate them. Therefore, the unattainability of QCRB not only impacts the precision, but also poses challenges in searching for optimal measurements~\cite{chen2023tradeoff,zhang2024qestoptpovm,PhysRevResearch.4.043057,yu2024quanestimationjl}.

The unattainability primarily arises from the incompatible nature of optimal measurements for different parameters, making their joint implementation unfeasible due to the Heisenberg uncertainty principle~\cite{OZAWA2004367}. 
The necessary and sufficient condition for saturating the multi-parameter QCRB is the weak commutativity condition~\cite{hayashi2008asymptotic,ragy2016compatibility,vidrighin2014joint,carollo2018uhlmann}. When this condition is satisfied, optimal measurements exist that achieve the ultimate limit simultaneously for different parameters. Hence, investigating which kinds of quantum states can satisfy this condition is essential in multi-parameter quantum metrology.
 Antiunitary symmetry, originating from the time-reversal symmetry, plays a crucial role in quantum mechanics~\cite{chaichian1998symmetries,uhlmann2016anti}. It has been shown that parameterized quantum
states with antiunitary symmetry always satisfy the weak commutativity condition~\cite{miyazaki2020symmetry}. In addition, this symmetry offers explicit methods to construct optimal measurements. Although antiunitary symmetry holds great potential in achieving ultimate precision, how to implement quantum metrology scheme
with antiunitary symmetric quantum states and perform
optimal compatible measurements practically remain to
be explored.

In this letter, we present the experimental demonstration showcasing the metrological advantages of antiunitary symmetry. By optimizing the encoding strategies, we design two types of antiunitary symmetric quantum statistical models, a set of parameterized quantum states with antiunitary symmetry. We implement the two models in optical systems and construct the optimal measurements via quantum walk~\cite{PhysRevLett.114.203602,hou_deterministic_2018}. Our findings demonstrate that the two models effectively eliminate the trade-off in the precision of different parameters, thereby resolving the inherent incompatibility and achieving the QCRB. Moreover, our results convincingly show the enhanced precision achieved in parameter estimation, thus paving a promising path for practical utilization of antiunitary symmetry in quantum metrology and sensing. 

\textit{Theoretical framework.}--The quantum statistical model of a physical system is denoted by $\{\rho_{\boldsymbol{\lambda}}\}$, where quantum states are parameterized by $n$ real parameters $\boldsymbol{\lambda}=(\lambda_1,\lambda_2,...,\lambda_n)^{\top}\in{\Lambda}\subseteq R^n$. The measurement is described by a positive operator valued measure (POVM) $\{\Pi_\chi |\Pi_\chi\geq 0,\sum_\chi\Pi_\chi=\mathbf{I}\}$ with outcomes $\{\chi\}$. According to Born's rule, the probability of obtaining outcome $ \chi$ conditioned on parameters $\boldsymbol{\lambda}$ is given by $p(\chi|\boldsymbol{\lambda})$=Tr$(\rho_{\boldsymbol{\lambda}}\Pi_\chi)$. The mean square error matrix (MSEm), denoted as $V_{\boldsymbol{\lambda}}$, quantifies the quality of an estimator $\boldsymbol{{\check \lambda}}$, and is defined as $V_{\boldsymbol{\lambda}}=\sum_\chi p(\chi|\boldsymbol{\lambda})(\boldsymbol{\check{\lambda}}(\chi)-\boldsymbol{\lambda})(\boldsymbol{\check{\lambda}}(\chi)-\boldsymbol{\lambda})^{\top}$, where diagonal elements represent variances in estimating different parameters and off-diagonal elements represent covariance between different parameters.
Precision bounds are given in terms of matrix inequalities for MSEm, Fisher information matrix (FIm) and quantum Fisher information matrix (QFIm),
\begin{figure}[t]
	\centering
	\includegraphics[width=\linewidth]{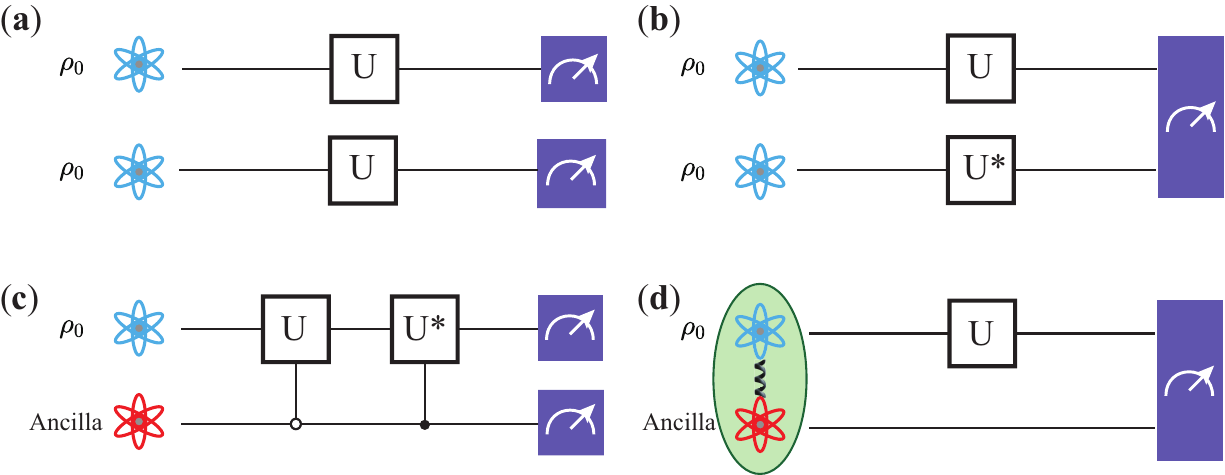}
	\caption{Schematic diagram illustrating strategies for estimating parameters encoded in the unitary operator U (blue squares). 
(a) Parallel model: Two identical quantum states undergo the same unitary process and are separately measured.
(b) Mutually conjugate model (MCM): Two identical quantum states undergo mutually conjugate unitary processes U and U* respectively and are measured jointly.
(c) Ancilla-assisted mutually conjugate model (AAMCM): One state controls another, determining whether it undergoes U or U*. The measurement is separable.
(d) Maximal entanglement model (MEM): Two maximally entangled states are generated, with one undergoing the unitary process U. Collective measurement is performed on them.}\label{figure1}
\end{figure}
\begin{equation}
    V_{\boldsymbol{\lambda}}\geq m^{-1}F^{-1}_{\boldsymbol{\lambda}} \geq m^{-1}Q^{-1}_{\boldsymbol{\lambda}}.
\end{equation}
Here, $m$ is the number of copies.
 The elements of FIm are given by $[F_{\boldsymbol{\lambda}}]_{ij}=\sum_{\chi} \frac{1}{p(\chi|\boldsymbol{\lambda})}\frac{\partial p(\chi|\boldsymbol{\lambda})}{\partial \lambda_i}\frac{\partial p(\chi|\boldsymbol{\lambda})}{\partial \lambda_j}$, and the elements of QFIm are given by $[Q_{\boldsymbol{\lambda}}]_{ij}$=$\frac{1}{2}$Tr$(\rho_{\boldsymbol{\lambda}}\{L_i,L_j\})$, where $\{\cdot\}$ is the anticommutator and $L_{i}$ is the symmetric logarithmic derivative (SLD), the solution of $\partial \rho_{\boldsymbol{\lambda}}/\partial \lambda_{i}=\frac{1}{2}(\rho_{\boldsymbol{\lambda}} L_{i}+L_{i}\rho_{\boldsymbol{\lambda}})$. There always exist estimators that make MSEm equal to the inverse of FIm \cite{kay1993fundamentals}. In the second inequality, the condition for the equality is that the mean Uhlmann curvature matrix $\mathcal{U}$ with the elements $\mathcal{U}_{ij}=\frac{i}{4}$Tr$(\rho_{\boldsymbol{\lambda}}[L_i,L_j])$  vanishes, the so-called weak commutativity condition \cite{hayashi2008asymptotic,ragy2016compatibility,vidrighin2014joint,carollo2018uhlmann}. To compare the precision represented by these matrices intuitively, the scalar precision bounds inequalities with a real symmetric weight matrix $\mathcal{W}$ are often utilized instead of matrix inequalities,
\begin{equation}\label{scalar}
    m\mathrm{Tr}(\mathcal{W}V_{\boldsymbol{\lambda}})\geq \mathrm{Tr}(\mathcal{W}F^{-1}_{\boldsymbol{\lambda}})\geq C^H_{\boldsymbol{\lambda}}(\rho_{\boldsymbol{\lambda}},\mathcal{W}) \geq \mathrm{Tr}(\mathcal{W}Q^{-1}_{\boldsymbol{\lambda}}),
\end{equation}
where $C^H_{\boldsymbol{\lambda}}(\rho_{\boldsymbol{\lambda}},\mathcal{W})$ is the Holevo Cramér-Rao bound (HCRB) \cite{holevo2011probabilistic,hayashi2005asymptotic,PhysRevLett.123.200503,PhysRevX.11.011028,demkowicz2020multi,guctua2006local,hayashi2008asymptotic,kahn2009local,yamagata2013quantum}, which represents the fundamental and asymptotically attainable bound with collective measurements on many identical copies of the quantum states. Especially for pure states, the HCRB is attained by separable measurements on single copies of $\rho_{\boldsymbol{\lambda}}$ \cite{Matsumoto_2002,conlon2021efficient}. The specific expression of the HCRB can be found in \cite{SM} and more details can be found in Ref.~\cite{holevo2011probabilistic,hayashi2005asymptotic,PhysRevLett.123.200503,PhysRevX.11.011028,demkowicz2020multi}. The HCRB equals the scalar QCRB only if the weak commutativity condition is satisfied.

\emph{Antiunitary symmetry}. The antiunitary operator $\Theta$ is unitary \big($\Theta^{\dagger}\Theta=\mathbf{I}$\big) and antilinear $\big(\Theta\sum_i\alpha_i|\phi_i\rangle=\sum_i\alpha_i^*\Theta|\phi_i\rangle,$ for any vectors $\{|\phi_i\rangle\}$ in Hilbert space, and complex coefficients $\{\alpha_i\}\big)$. An antiunitary and Hermitian operator $\vartheta$ is called a $conjugation$ operator, i.e. $\vartheta\vartheta^{\dagger}=\mathbf{I}$ and $\vartheta=\vartheta^{\dagger}$. Conjugation operators can transfer Hermitian operators into their complex conjugation, $\vartheta \rho_{\boldsymbol{\lambda}} \vartheta =\rho_{\boldsymbol{\lambda}}^*$. 

A class of quantum statistical models with antiunitary symmetry is defined as follows~\cite{miyazaki2020symmetry}: 
If there exists an antiunitary operator $\Theta$ such that $\Theta\rho_{\boldsymbol{\lambda}}\Theta^{\dagger}=\rho_{\boldsymbol{\lambda}}$ holds for any $\boldsymbol{\lambda}$, $\{\rho_{\boldsymbol{\lambda}}\}$ is said to have \emph{global antiunitary symmetry} (GAS). An important property of the quantum state $\rho_{\boldsymbol{\lambda}}$ with GAS is that it satisfies the weak commutativity condition at all points in the parameter space,
\begin{equation}\label{wcc}
    \mathcal{U}_{i j}(\rho_{\boldsymbol{\lambda}})=0,  \forall i,j.
\end{equation}

\emph{Mutually conjugate model}. One kind of quantum statistical model with GAS is 
\begin{equation}\label{3}
    \{\rho_{1}={\rho_{\boldsymbol{\lambda}} \otimes \rho_{\boldsymbol{\lambda}}^*}|\lambda\in \Lambda\},
\end{equation}
with antiunitary operator $S\vartheta^{\otimes 2}$, where $S$ is the swap operator of the bipartite system. Here, we call it \emph{mutually conjugate model} (MCM). The QFIm of the quantum states in the MCM is twice that of the original quantum state $\rho_{\boldsymbol{\lambda}}$ and is equal to that of the parallel model $\{\rho_{\boldsymbol{\lambda}} \otimes \rho_{\boldsymbol{\lambda}}|\lambda\in \Lambda\}$, while the latter cannot always achieve the QCRB.

\begin{figure*}[t]
	\centering
	\includegraphics[width=\linewidth]{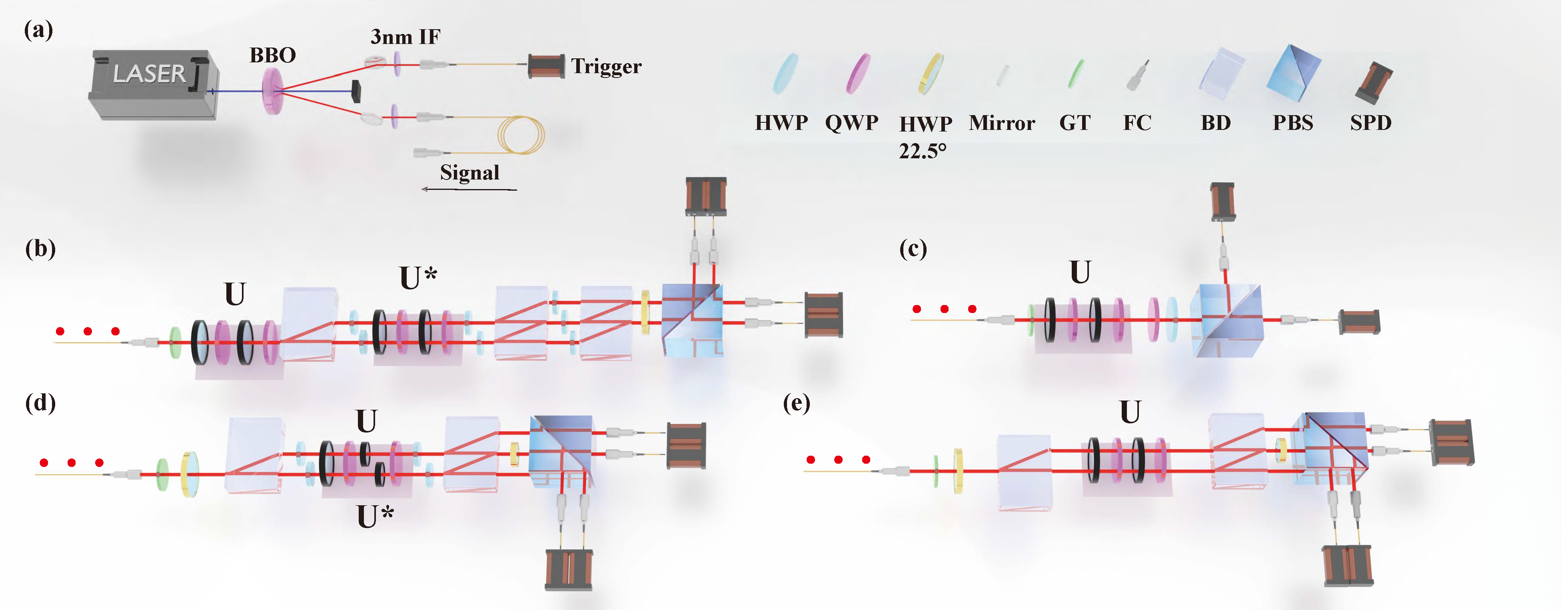}
	\caption{The experimental setup for proving the metrological advantages of quantum states with GAS. There are five modules in the setup. $\textbf{(a)}$ The single-photon source module. $\textbf{(b)}$ The MCM. $\textbf{(c)}$ The parallel model. $\textbf{(d)}$ The AAMCM. $\textbf{(e)}$ The MEM. BBO $\beta$-barium borate crystal, IF interference filter, GT Glan-Thompson, FC fiber coupler, BD beam displacer, PBS polarization beam splitter, SPD single photon detector.}\label{figure2}
\end{figure*}

We consider the problem of estimating unknown parameters ($\theta,\phi$) in qubits, as represented by,
\begin{equation}\label{qubit}
|\psi_{{\boldsymbol{\lambda}}}\rangle=\cos\frac{\theta}{2}|0\rangle+e^{i\phi}\sin\frac{\theta}{2}|1\rangle,
\end{equation}
with $\theta\in[0, \pi]$, $\phi \in[0,2 \pi]$. The QFIm with respect to $(\theta,\phi)$ in Eq.~(\ref{qubit}) is a diagonal matrix, $Q_{{\boldsymbol{\lambda}}}$=diag(1,$\sin^2 \theta$). We choose the cost matrix $\mathcal{W}$=diag$(1,\sin^2 \theta)$ which is related to the natural metric on the sphere, also known as the Fubini-Study metric \cite{bengtsson2017geometry}. The scalar QCRB gives $\operatorname{Tr}\left(\mathcal{W} Q_{\boldsymbol{\lambda}}^{-1}\right)$=2. However, the weak commutativity condition between parameters $(\theta,\phi)$ is not satisfied, as $\mathcal{U}_{\theta\phi}\left(|\psi_{\boldsymbol{\lambda}}\rangle\right)\neq0$, therefore the scalar QCRB cannot be achieved. Meanwhile, the HCRB $C^H_{\boldsymbol{\lambda}}(|\psi_{_{\boldsymbol{\lambda}}}\rangle,\mathcal{W})=4$. We see that HCRB is twice the scalar QCRB, which corresponds to the maximal discrepancy \cite{Carollo_2019,tsang2020quantum}. For pure states, the HCRB can be achieved by separable measurements performed on each copy of $|\psi_{{\boldsymbol{\lambda}}}\rangle$. Due to the fact that $C^H_{\boldsymbol{\lambda}}(|\psi_{_{\boldsymbol{\lambda}}}\rangle,\mathcal{W})=n C^H_{\boldsymbol{\lambda}}({|\psi_{_{\boldsymbol{\lambda}}}\rangle}^{\otimes n},\mathcal{W})$~\cite{demkowicz2020multi}, the precision of parallel models $|\psi_{\boldsymbol{\lambda}}\rangle^{\otimes 2}$ is twice that of single copies, $C^H_{\boldsymbol{\lambda}}(|\psi_{{\boldsymbol{\lambda}}}\rangle^{\otimes 2},\mathcal{W})=2$, and the scalar QCRB of $|\psi_{{\boldsymbol{\lambda}}}\rangle^{\otimes 2}$ can not be achievable.

For the MCM, the weak commutativity condition always holds. The scalar QCRB of the MCM
\begin{equation}
\begin{aligned}
    C^Q_{\boldsymbol{\lambda}}(\left|\psi_{\boldsymbol{\lambda}}\right\rangle \otimes \left|\psi^*_{\boldsymbol{\lambda}}\right\rangle,\mathcal{W})=\frac{1}{2}C^Q_{\boldsymbol{\lambda}}(\left|\psi_{\boldsymbol{\lambda}}\right\rangle,\mathcal{W})&\\=\frac{1}{4}C^H_{\boldsymbol{\lambda}}(|\psi_{\boldsymbol{\lambda}}\rangle,\mathcal{W})=\frac{1}{2}C_{\boldsymbol{\lambda}}^{H}\left(\left|\psi_{\boldsymbol{\lambda}}\right\rangle^{\otimes 2}, \mathcal{W}\right),
\end{aligned}
\end{equation}
where $C^Q_{{\boldsymbol{\lambda}}}(\star,\mathcal{W})=\operatorname{Tr}\left(\mathcal{W} Q_{{\boldsymbol{\lambda}}}^{-1}\right)$.

\emph{Ancilla-assisted mutually conjugate model}.
A quantum statistical model $\{|\Psi_{\boldsymbol{\lambda}}\rangle_A\langle \Psi_{\boldsymbol{\lambda}}||\boldsymbol{\lambda}\in\Lambda\}$, where
\begin{equation}\label{EQS}
|\Psi_{\boldsymbol{\lambda}}\rangle_A=\frac{|\psi_{\boldsymbol{\lambda}}\rangle|0\rangle+|\psi^*_{\boldsymbol{\lambda}}\rangle|1\rangle}{\sqrt{2}},
\end{equation}
also has GAS with an antiunitary operator $(I\otimes\sigma_x) \vartheta$. We call this model \emph{ancilla-assisted mutually conjugate model} (AAMCM).
The QFIm of $|\Psi_{\boldsymbol{\lambda}}\rangle_A$ is
\begin{equation}
       Q_{A}= \begin{bmatrix}
    1&0\\0&2-2\cos{\theta}
    \end{bmatrix}.
\end{equation}
To highlight the advantage of the AAMCM, we take another ancilla-assisted quantum statistical model---the maximal entanglement model (MEM) $\{|\Psi_{\boldsymbol{\lambda}}\rangle_{MEM}\langle \Psi_{\boldsymbol{\lambda}}||\boldsymbol{\lambda}\in\Lambda\}$ for comparison, in which
\begin{equation}  
\begin{aligned}
|\Psi_{\boldsymbol{\lambda}}\rangle_{MEM}&=\frac{|\psi_{\boldsymbol{\lambda}}\rangle|0\rangle+|\psi_{\boldsymbol{\lambda}}^{\perp}\rangle|1\rangle}{\sqrt{2}},
\end{aligned}
\end{equation}
with $\langle \psi_{\boldsymbol{\lambda}}|\psi_{\boldsymbol{\lambda}}^{\perp}\rangle=0$. 
The MEM also has the GAS with the antiunitary operator $\Theta_f\otimes\Theta_f$, where $\Theta_f$ is the bit flip operator \cite{miyazaki2020symmetry}. The MEM is viewed as the most sensitive quantum statistical model to estimate the parameters encoded in it \cite{PhysRevLett.87.177901,PhysRevLett.117.160801}, when the encoding process is unitary. The AAMCM ensures the same level of precision in parameter estimation as the MEM.

\textit{Measurement.}--
The existence of a set of optimal POVMs that saturate the scalar QCRB for pure states with GAS is always guaranteed. The work in Ref.~\cite{miyazaki2020symmetry} presents a method to identify parameter-independent optimal POVMs. If the quantum state can be decomposed into parameter-independent eigenvectors with real coefficients, these eigenvectors also form the optimal POVM. The optimal measurements for MCM and AAMCM are provided below.

The elements of the optimal POVM for the MCM are 
\begin{equation}\label{POVM_MCM}
\begin{aligned}
    &\Pi_1=|\Phi^+\rangle\langle\Phi^+|,\ \Pi_2=|\Phi^-\rangle\langle\Phi^-|, \\&\Pi_3=|\Psi^+\rangle\langle\Psi^+|,\ \Pi_4=|\Psi^-\rangle\langle\Psi^-|,
\end{aligned}
\end{equation}
where $|\Phi^{\pm}\rangle=1/\sqrt{2}(|00\rangle\pm|11\rangle)$ and $|\Psi^\pm\rangle=1/\sqrt{2}(|01\rangle\pm|10\rangle)$. The elements of the optimal POVM for the AAMCM are 
\begin{equation}\label{POVM_AAMCM}
\begin{aligned}
    &\pi_1=|0\rangle\langle 0|\otimes|0\rangle\langle 0|,\pi_2=|0\rangle\langle 0|\otimes|1\rangle\langle 1|,\\
&\pi_3=|1\rangle\langle1|\otimes|+\rangle\langle+|,\pi_4=|1\rangle\langle1|\otimes|-\rangle\langle-|,
\end{aligned}
\end{equation}    
where $|+\rangle=(|0\rangle+|1\rangle)/\sqrt{2}$ and $|-\rangle=(|0\rangle-|1\rangle)/\sqrt{2}$. From the Eq.~(\ref{POVM_AAMCM}), we can find the optimal measurement for the AAMCM is separable. The MEM also has the GAS, we can also give the optimal measurement for it,
\begin{equation}\label{POVM_MEM}
\begin{aligned}
&\Pi_1^U=|00\rangle\langle00|,\Pi_2^U=|11\rangle\langle11|,\\
&\Pi_3^U=|\Psi^+\rangle\langle\Psi^+|,\Pi_4^U=|\Psi^-\rangle\langle\Psi^-|.
\end{aligned}
\end{equation}
The measurement proposed in Ref.~\cite{PhysRevLett.117.160801} consists of four Bell projectors, while the POVM in Eq.~(\ref{POVM_MEM}) only includes two entangled projectors, namely $\Pi_3^U$ and $\Pi_4^U$. 

\textit{Experimental setup and results.}-- 
The experiment setup for proving the metrological advantages of GAS is presented in Fig.~\ref{figure2}. The experiment setup consists of five parts: a single-photon source and four different models - the MCM, the parallel model, the AAMCM, and the MEM.

In the single-photon source part, a pulse laser at 415 nm is input to a beta barium borate ($\beta$-BBO) crystal. This crystal is used for type-II beamlike spontaneous parametric downconversion (SPDC) to produce degenerate photon pairs with a central wavelength of 830 nm. The photon pairs are spectrally filtered by interference filters with 3 nm full-width at half-maximum and detected by single-photon detectors (SPDs). The single photon is heralded by detecting the other one. 

We prepare the MCM in Fig.~\ref{figure2} (b). The first qubit is encoded in the polarization degree of freedom (DOF) of the photon, where $|0\rangle$ and $|1\rangle$ correspond to horizontal and vertical polarization respectively. The second qubit is encoded in the path DOF of the photon, with $|up\rangle$ and $|down\rangle$ representing $|0\rangle$ and $|1\rangle$. Arbitrary qubit states are generated using a combination of wave plates consisting of two half-wave plates (HWPs) and two quarter-wave plates (QWPs). One HWP controls the parameter $\theta$, while another one controls $\phi$. Both QWPs are fixed at a rotation angle of 45°. A beam displacer (BD) is utilized to displace components with horizontal polarization into the up path, and those with vertical polarization into the down path, thereby mapping the polarization qubit into the path DOF. On each path, another set of wave plates is employed to encode parameters $\theta$ and $-\phi$ into the polarization DOF of photons. This enables us to prepare an MCM state given by $\left|\psi_{\boldsymbol{\lambda}}\right\rangle \otimes\left|\psi_{\boldsymbol{\lambda}}^{*}\right\rangle$. The measurement component realizes collective measurement as described in Eq.~(\ref{POVM_MCM}). We implement this POVM through quantum walk involving several BDs and HWPs. Our method utilizes two DOFs of single photons to generate 2-qubit quantum states and realizes collective measurements deterministically on the states. This method is more reliable and has higher fidelity compared with the 2-photon systems~\cite{PhysRevLett.118.050501,RN39,PhysRevLett.96.190501}, which is crucial to achieve the optimal precision.
\begin{figure}[t]
	\centering
	\includegraphics[width=\linewidth]{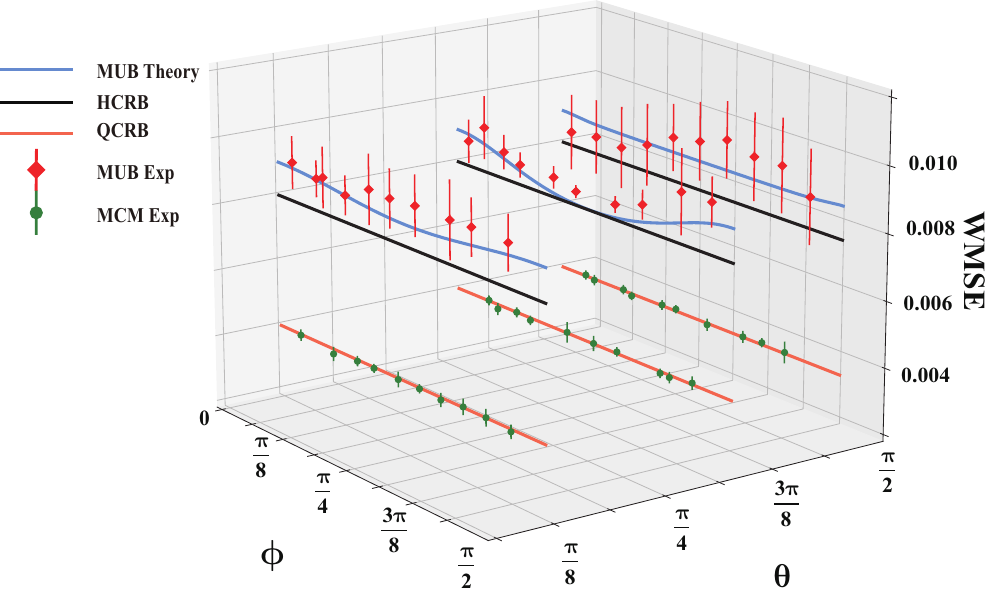}
	\caption{Experimental results for estimating $(\theta,\phi)$ simultaneously of the MCM and parallel model. We select three groups of quantum states with $\theta$=0.4,1,1.4 rad respectively. The number of total qubits $m$ is 500.}\label{result}
\end{figure}
In Fig.~\ref{figure2}(c), we prepare $|\psi_{\boldsymbol{\lambda}}\rangle$ encoded in the polarization DOF of the photon by HWPs and QWPs. We realize the mutually unbiased bases-like (MUB-like) POVM to perform projection on certain basis by utilization of HWPs, QWPs followed by a polarization beam splitter. The HCRB of the parallel model is achievable by the  MUB-like POVM solely on a single copy twice. 
 
The experimental setup of the AAMCM is shown in Fig.~\ref{figure2}(d). The quantum state is encoded in the polarization, and the path DOF acts as the control bit to prepare  $(|\psi_{\boldsymbol{\lambda}}\rangle|0\rangle+|\psi^*_{\boldsymbol{\lambda}}\rangle|1\rangle)/\sqrt{2}$. Similar to MCM, two HWPs are used to put $|10\rangle$ and $|11\rangle$ of the quantum state into the same path and they would interfere at an HWP with a rotation angle 22.5$^{\circ}$ to realize the measurement in Eq.~(\ref{POVM_AAMCM}). 

The MEM is shown in Fig.~\ref{figure2}(e). After passing through an HWP and a BD, the photon is prepared in the state $(|H,up\rangle+|V,down\rangle)/\sqrt{2}$. The polarization qubit will be manipulated by the group of wave plates to generate the quantum state in Eq~(\ref{qubit}) and the path DOF acts as an ancillary qubit. The POVM in Eq.~(\ref{POVM_MEM}) is realized by a BD and a HWP with a rotation angle 22.5$^{\circ}$.

The measurement results of all of the models are recorded by single-photon detectors. The experimental results of the MCM are shown in Fig.~\ref{result}. To estimate $\theta$ and $\phi$ based on the statistics of measurement outcomes, we employ the Bayesian estimation method with a uniform prior probability distribution for $\theta$ and $\phi$. We update the posterior probability distribution after each sequence $m$ of experimental measurement results.  Due to periodicity in the probability distributions of the measurement outcomes, as well as potential existence of identical measurement results for different parameters, we constrain the parameters ($\theta$,\ $\phi$) to lie between 0 and $\pi$/2 to ensure a one-to-one correspondence between estimators and measurement results. The estimate and uncertainty of the parameter are thus assessed by calculating the first and second moments of the posterior probability. We select three sets of quantum states with $\theta$=0.4, 1, 1.4 rad respectively, and each set of quantum states contains 10 different values of $\phi$. The figure shows that the experimental results with 500 qubits per trial of weighted mean squared error $\operatorname{Tr}\left(\mathcal{W} V_{\boldsymbol{\lambda}}\right)$ for the MCM match the theoretical predictions. Furthermore, the measurement we used for the MCM is optimal for all parameters, whereas the MUB-like POVM performed on the parallel model cannot always saturate the HCRB for each parameter. 

\begin{figure}[t]
\centering
\includegraphics[width=\linewidth]{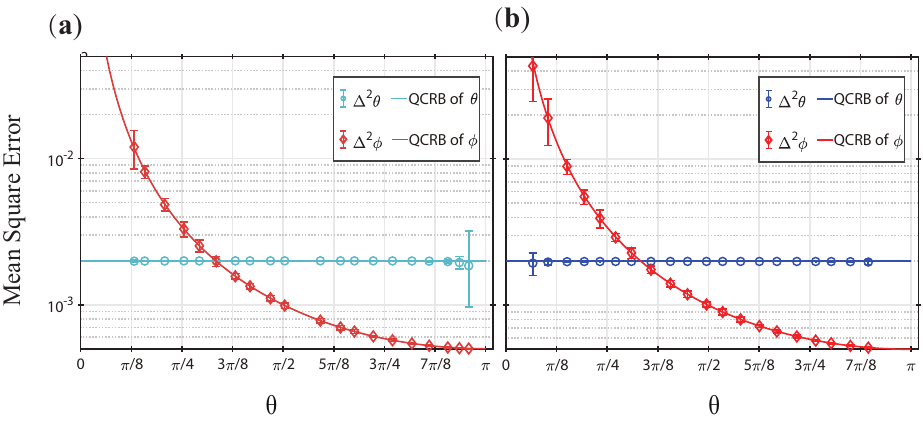}
\caption{Experimental results for estimating $(\theta,\phi)$ simultaneously of the AAMCM and MEM. In the experiment, the parameter $\phi$=$\pi$/4 and we change different $\theta$. $\textbf{(a)}$ Experimental results of the AAMCM. $\textbf{(b)}$ Experimental results of MEM. The number of total qubits $m$ is 500.}\label{result2}
\end{figure}

The experimental results of the AAMCM and the MEM are shown in Fig.~\ref{result2}. The parameter $\phi$ is fixed as $\pi/4$, and $\theta$ is scanned from 0 to $\pi$. It can be seen the experimental mean square errors match the theoretical lines. The AAMCM can achieve the same maximal precision as the MEM.

\textit{Summary and discussions.}--In summary, we experimentally demonstrate the metrological advantage of a class of quantum states with GAS. 
 We produce two types of quantum states with GAS in a photonic platform, i.e. the MCM and the AAMCM. The MCM in Eq.~(\ref{3}) is a simple example with GAS that achieves better precision than two identical states.
Measurement of classically correlated quantum states with collective measurements is related to the phenomenon called nonlocality without entanglement~\cite{PhysRevA.59.1070} , and has been demonstrated to improve the accuracy of quantum state discrimination~\cite{PhysRevLett.94.220406}. Both the discrimination and metrological advantages demonstrated here are due to the global geometry structures of classically correlated quantum states, i.e. the MCM in our work, which can be more efficiently extracted by collective measurement. 
Meanwhile, we experimentally demonstrate the AAMCM, another quantum model with GAS, achieves the same maximal precision as the maximally entangled states but only needs separable measurement.

Even though the MCM and AAMCM have great advantages in multi-parameter quantum metrology, implementing the conjugation of an unknown unitary evolution is challenging. For a real Hamiltonian ($\mathbf{H}^*=\mathbf{H}$), conjugating the unitary operation $\mathbf{U}=\exp(-i\mathbf{H}t)$ corresponds to its inverse operation $\mathbf{U}^{-1}=\exp(i\mathbf{H}t)$. However, it is impossible to deterministically and exactly implement the inverse operation $\mathbf{U}^{-1}$ with just a single use of $\mathbf{U}$ due to a no-go theorem \cite{Chiribella_2016,PhysRevLett.122.170502,PhysRevA.100.062339,PhysRevLett.123.210502}. Nevertheless, for qubit systems, the universally time-reversal or inverse operation can be realized deterministically and exactly \cite{PhysRevLett.131.120602,Schiansky:23,doi:10.1126/sciadv.1602589,stromberg2022experimental}. Our work opens up a new direction for exploring the metrological power of a class of quantum statistical models with GAS. It is anticipated to find more quantum statistical models with GAS showcasing their advantages in quantum information processing, not only for exploring the metrological advantages but also showing the quantum superiority in quantum state estimation \cite{chang2014optimal,tang2020experimental} and quantum communication \cite{gisin1999spin}.

\begin{acknowledgments}
This work was supported by the National Key Research and Development Program of China (Grants No. 2023YFC2205802, 2019YFA0308700), National Natural Science Foundation of China (Grants No. 12347104, 12305034), Project funded by China Postdoctoral Science Foundation (2022M721565) and Civil Aerospace Technology Research Project (D050105).
\end{acknowledgments}

\end{document}